\documentclass[usenatbib, usegraphicx]{mn2e}

\newcommand{\apj}{\textit{ApJ}}
\newcommand{\aj}{\textit{AJ}}

\newcommand{\apjl}{\textit{ApJ Lett.}}
\newcommand{\apjs}{\textit{ApJ Suppl.}}

\newcommand{\mnras}{\textit{MNRAS}}

\newcommand{\BoxS}{S$_1$}
\newcommand{\BoxCR}{S$_2$}
\newcommand{\BoxCRgad}{S$_3$}
\newcommand{\msun}{M_\odot}

\def\disp{\displaystyle}

\title[]{The emptiness of voids: yet another over-abundance problem for the LCDM model.}
\author[A. Tikhonov and
A.  Klypin]{Anton V. Tikhonov$^{1}$\thanks{E-mail: ti@hotbox.ru;
    avt@gtn.ru } and Anatoly Klypin$^{2}$\thanks{E-mail:
    aklypin@nmsu.edu }\\
  $^{1}$Universitetsky prospect, 28, Department of Mathematics  and
  Mechanics, St.Petersburg State University, \\ Saint-Petersburg,
   198504, Russian Federation\\
  $^{2}$Department of Astronomy, New Mexico State University, Las
  Cruces, New Mexico 88003-8001, USA}
\begin{document}

\date{Accepted 2008 February ??. Received ?? February ??; in original form 2008 February ??}

\pagerange{\pageref{firstpage}--\pageref{lastpage}} \pubyear{2002}

\maketitle

\label{firstpage}

\begin{abstract}
  We study the luminosity function, the peculiar velocities, and the
  sizes of voids in the Local Volume (LV) in observational samples of
  galaxies which contain galaxies down to $M_B=-10$ and to $M_B=-12$
  within the distance $4- 8$~Mpc.  When we compare the results with
  the predictions of the standard cosmological LCDM model, we find
  that the theory faces a sever problem: it predicts a factor of ten
  more dwarf haloes as compared with the observed number of dwarf
  galaxies.  In the LV we identify voids with sizes ranging from 1 to
  4.5\,Mpc and compare the observational distribution of 
  void sizes with the voids in very high resolution simulations of the LCDM
  model with WMAP1 and WMAP3 parameters. The theoretical void function
  matches the observations remarkably well {\it only} if we use haloes
  with circular velocities $V_c$ larger than $40-45$\,km/s ($M_{\rm
    vir}=(1-2)\times 10^{10}\msun$) for models with with $\sigma_8 =
  0.9$ and $V_c>35$\,km/s ($M_{\rm vir}=(6-8)\times 10^{9}\msun$) for
  $\sigma_8 = 0.75$.  We exclude the possibility that in the LCDM
  model haloes with circular velocities $< 35$\,km/s can host galaxies
  as bright as $M_B=-12$: there are too many small haloes in the LCDM
  model resulting in voids  being too small as compared with
  the observations.  The problem is that many of the observed dwarf galaxies
  have HI rotational velocities below 25\,km/s that
  strictly contradicts the LCDM predictions. Thus, the LCDM model
  faces the same overabundance problem, which it had with the number
  of satellites in the LG.  We also estimate the rms deviations from
  the Hubble flow $\sigma_H$ for galaxies at different distances from
  the Local Group and find that in most of our model LV-candidates the
  rms peculiar velocities are consistent with observational values:
  $\sigma_H=50$\,km/s for distances less than 3\,Mpc and
  $\sigma_H=80$\,km/s for distances less than 8\,Mpc. At distances 4
  (8) Mpc, the observed overdensities of galaxies are 3.5-5.5
  (1.3-1.6) -- significantly larger than typically assumed.
\end{abstract}


\section{Introduction}\label{sec:intro}
Voids in the distribution of galaxies are an important ingradient of the
distribution of light and mass in the Universe.  They constitute a
natural outcome of structure formation via gravitational instability,
and, thus, can be used to constrain theories of galaxy formation.
Emptiness of voids -- the number of small galaxies in the voids -- is
an interesting question for both the observations and the theory to
tackle
\citep{Peebles2001,Gottloeber2003,Patiri2006,Tinker08}. Cosmological
simulations predict \citep[e.g.,][]{Gottloeber2003} that many small DM
haloes should reside in voids. There seems to be no disagreement
between the $\Lambda$CDM theory and the observations regarding the
giant voids defined by $M_*$ galaxies or by $\sim 10^{12}M_{\odot}$ haloes
\citep{Patiri2006}. The situation is less clear on smaller scales.  In
the region of $\sim 10$\,Mpc around the Milky Way, where observations
go to remarkably low luminosities, small voids look very empty: dwarf
galaxies do not show a tendency to fill the voids and voids are still
relatively large. The theory predicts that many dwarf dark matter
haloes should be in the voids, which puts it on a collision course with
observations. Yet, below some mass the haloes are expected to stop
producing galaxies inside them. There are different arguments for
that: stellar feedback \citep{Dekel1986} or photoionization
\citep{Bullock00} may play a significant role in quenching star
formation in too small haloes. For example, \citet{Loeb08} made a
simple estimation of the limiting circular velocity below which haloes
have essentially no gas infall due to increase of Jeans mass caused by
UV background at the epoch of reionization: $V_{lim} = 34 \times
(T_{IGM}/1.5\times10^4K)^{1/2}$\,km/s, where $T_{IGM}$ is the
temperature of intergalactic medium gas ionized by stars.
\citet{Hoeft06} studied formation of dwarf DM haloes in cosmological
void regions using high-resolution hydrodynamic simulations and
assuming that cosmological UV-background photo-evaporates baryons out
of haloes of dwarf galaxies, and thereby limits their cooling and star
formation rate.  \citet{Hoeft06} give characteristic mass $M_c = 6
\times 10^9 h^{-1}M_\odot$ below which haloes start to fail accreting
gas.

Theoretical estimates for the least massive luminous halo are still
uncertain.  It is difficult to get a definite answer because the
physics of  dwarf galaxies at high redshifts is quite
complicated. Star formation histories of some isolated irregular
dwarfs indicate that starbursts may produce enough power to throw gas
away, but not sufficient for galaxy to get rid of it -- gas again
returns to potential well of the DM halo, hosting the galaxy
\citep{Quillen08}.

Satellites of the Local Group shed light on the problem from a
different direction.  The $\Lambda$CDM model predicts that thousands
of dwarf DM haloes should exist in the Local Group \citep{Klypin99,
  Moore99, Madau08}, while only $\sim 50$ are observed. Recent
discoveries of very low luminosity dwarfs \citep{Simon07} and careful
analysis of incompleteness effects in SDSS \citep{Strigari07,Simon07,Tollerud08}
bring the theory and observations a bit closer, but the mismatch still seems
to be present.  The currently favored  explanation of the
overabundance of the dark matter subhaloes
\citep{Bullock00,Kravtsov2004} assumes that dwarf haloes above $V_c
\approx 30-50$\,km/s were forming stars before they fall into the
Milky Way or M31 and that smaller haloes never formed any substantial
amount of stars. Once the haloes above the limit fall into the halo of
the Milky Way or M31, they get severely stripped and may substantially
reduce their circular velocity producing galaxies such as Draco or
Fornax with the rms line-of-sight velocities onlya few km/s. The
largest subhaloes retain their gas and continue to form stars, while
smaller ones may lose the gas and become dwarf spheroidals. Haloes
below the limit never had substantial star formation. They are truly
dark. This scenario implies that circular velocity before the infall
$V_c\approx 30-50$\,km/s is the limit for star formation in
haloes. \citet{Moore06} give additional arguments in favor of this
scenario.

If this picture is correct, it can be tested using the abundance and
the distribution of dwarf galaxies outside of the Local Group. Because
dwarfs can only be detected at small distances, useful observational
samples are limited to distances less than 10\,Mpc.  

While the Local Volume sample of galaxies is not as deep as the sample
of satellite galaxies in the Local Group, we argue that the Local
Volume dwarfs provide a unique opportunity to study the smallest and
the darkest galaxies. The problem with using the LG satellites is related with
the fact that the satellites have been tidally stripped by the Milky Way or by M31. Indeed,
theoretical estimates indicate that a very substantial mass loss occurs even
at very small distances from the centres of the dwarfs
\citep[e.g.,][]{Madau08}. Thus, we really do not know whether we
deal with a truly low mass and low circular velocity satellite or with a
satellite, which was much more massive in the past and was later
severely stripped by its parent galaxy. The Local Volume dwarfs
represent a more pristine sample in this respect.

The problem of the emptiness of voids was recently re-visited by
\citet{Tinker08} with the conclusion that voids are not a problem for
the $\Lambda$CDM model.  Most of the observational results used in
\citet{Tinker08} are based on the void probability function in the
SDSS DR6 sample and on the nearest neighbor statistics in the ORS
catalog as analyzed by \citet{Peebles2001}.  The void probability
function was estimated for relatively bright galaxies with
$M_r<-17$. The nearest neighbor statistics goes slightly deeper: the
ORS sample is formally complete to $m_B=14.5$, which gives $M_B\approx
-15$ at the distance of 8\,Mpc. Our sample is 3 magnitudes deeper,
which is essential for the ``void phenomenon''. We also note that the
quality of our sample is much better than that of the ORS sample. For
example, we use real distances to galaxies, not redshifts. Special
effort was made to ensure that the sample is complete for LSB galaxies
and for galaxies in the zone of avoidance. 

\citet{Tikhonov2006} presented results on statistics of nearby voids
in the Local Volume. Here we continue the analysis using an updated
list of galaxies (Karachentsev, private communication).  We
characterize the spatial distribution of galaxies in the LV mostly by
studying the distribution of sizes of  quasi-spherical
regions - voids. The voids may still contain gas and small dark matter
haloes. For our purpose void statistic is reasonably robust since 3d
void maps are not very sensitive to a total number of objects in a
sample. The distribution of void sizes have the advantage that they
are sensitive to appearance of galaxies in very low density
environments: an important property for studying the smallest existing
galaxies.

We compare the spectrum of void sizes in the LV with the distribution
of voids in high-resolution cosmological simulations. The simulations
give us detailed information (positions, velocities, masses, circular
velocities, and so on) for dark matter haloes and their
satellites. However, they do not provide luminosities of galaxies.
Theoretical predictions of the luminosity of a galaxy hosted by a halo
with given mass, circular velocity and merging history are quite
uncertain and cannot be used for our analysis. Instead, we ask a more
simple question: what luminosity a halo or subhalo with given circular
velocity {\it should have} in order to reproduce the observed spectrum
of void sizes. When doing this, we assume that haloes with larger
circular velocities should host more luminous galaxies. We will later
see that matching of the void spectrum in simulations and with the
observations puts significant constraint on relation of the halo
circular velocity and the luminosity of a galaxy hosted by the
halo. If we take too large circular velocity, there are too few
galaxies and sizes of voids become too large. Instead, if very small
haloes host galaxies, the number of large voids declines well below
what is observed in the Local Volume.

In addition to the analysis of distribution of voids in the Local Volume we
also re-visit the problem of the deviations from the Hubble flow.  
The flow of field galaxies in LV appeared
to be rather ``cold'': deviations from the Hubble velocity are rather
small.  For example,  using the
Tully-Fisher distances and applying error correction via quadrature
subtraction \citet{Shlegel94} have found for 15 galaxies within
500~km/s and outside Local Group the rms peculiar velocity
$\sim$60~km/s. They noted that such
values are very rare in the CDM models, but are not uncommon in MDM
models that include massive neutrinos.  Comparable values have been
derived by \citet{KM96} for galaxies within 7~Mpc from the Local
Group.  \citet{Maccio05} based on their galaxy sample obtained
$\sigma_H \sim 52$\,km/s within 3\,Mpc.

\citet{KM01} found evidence of anisotropy of Hubble flow in LV. There
is a dipole component, which is due to the LG motion relative to
galaxies in the LV. This is typically described as apex motion of the
LV.  There is also a quadrupole component of the deviations from the
Hubble flow interpreted as anisotropic expansion of Local
Volume. \citet{KM01} removed the apex motion and the quadrupole
component from estimates of $\sigma_H$.  They also removed galaxies
with large velocity deviations, which were considered due to infall of
galaxies onto large groups inside the LV. Galaxies inside groups
were also not considered.
Using only the galaxies with accurate measurements of distances  and assuming
that errors in distances increase the rms  peculiar velocities,
\citet{KM01} estimate $\sigma_H$ in distance range 1-3 Mpc from the
centre of LG may be as low as $\sim 30$ km/s. \citet{Kar03} found
$\sigma_H \sim 40$\,km/s inside a sphere of 5~Mpc using distances from
the luminosity of the tip of the red giant branch stars of about 20 dwarf
galaxies. 
Their procedure of $\sigma_H$ estimation include determination of the
``local'' Hubble constant, which is slightly different from universal
expansion rate: the true Hubble constant. 

All these effects and corrections are valid, but they have a tendency
to systematically underestimate the deviations from the Hubble
flow. Some of the corrections can be mimiced in simulations, but it is
difficult to do all of them. At the same time, one should not use
these corrections: any deviation from the global Hubble flow must be
included in the estimates. We make only one exclusion. We still
consider the apex motion because it is related with the selection of
the reference frame in which the whole sample does not have a net
velocity.

There were several attempts to study $\sigma_H$ in simulations.
\citet{Gov97} emphasized  that the dispersion of random motions of
field galaxies and the centres of groups allow to discriminate
between models with different values of matter density $\Omega_M$.
\citet{Klypin03} obtained in their constrained ``Local
Supercluster'' simulation the peculiar line-of-sight velocity
dispersion within $7h^{-1}$\,Mpc of the model LG  $\sigma_H\sim
60$\,km/sec comparable to the observed velocity dispersion of
nearby galaxies. They emphasize that there is no need in exotic
explanations of the ``coldness'' of Hubble flow. \citet{Maccio05}
 have found that in their simulations the Hubble
flow is significantly colder around model LG-candidates selected
in a $\Lambda$CDM cosmology than around LG-candidates in open or
critical models. Their estimation of $\sigma_H$ was much simpler
then \citet{KM01} approach -- they calculate rms around mean value
of peculiar velocity with local Hubble constant (best fit to
data). 

A key question for our kind of investigation is to define what an
``LV-candidate'' is.  In other words, it is important to define our local
environment. Usually the conditions for selection of LG-candidates are
rather simple and include distances of the LG-candidate to the nearest
Virgo-type cluster \citep[e.g.,][]{Hoffman08} and the overdensity of a
sphere of 8~Mpc radius \citep{Maccio05}.  \citet{Shlegel94} used the
overdensity 0.25. \citet{Maccio05} used $0.1 < \delta\rho / \rho <
0.6$. \citet{Karachentsev2004} noted that inside the radius of 8\,Mpc
centred on the Local Group the luminosity density in the B-band is
1.8 -- 2 times larger than the average luminosity density. 

\section{DATA: Local Volume}\label{sec:data}
Over the past few years searches for galaxies with distances less than
10\,Mpc have been undertaken using numerous observational data
including searches for LSB galaxies, blind HI surveys, and NIR and HI
observations of galaxies in the zone of avoidance
\citep{Karachentsev2004,Kar07ap}. At present, the sample contains
about 550 galaxies.  The distances to the galaxies are not measured
using the redshifts because the perturbations of the Hubble flow in
the Local Volume are large and significantly distort the spatial
distribution of galaxies. The distances are mostly measured with the
tip of the red giant brunch (TRGB) stars, cepheids, the Tully-Fisher
relation, and some other secondary distance indicators.  For most of
galaxies the distances have been measured with the accuracy of
8-10\% \citep{Karachentsev2004}.

\begin{table}
\caption{Test of sample completeness: Counts of galaxies with different absolute magnitudes $M_B$ in radial shells.}
\begin{tabular}{cccl}
\hline
 Radial bin      & $M_B<-15$ & $M_B=-12-14.5$  &  Ratio  \\
 (Mpc)      &   \\
\hline
1 -- 2 & 1 &  3&  3 \\
2 -- 3 & 10 & 8 & 0.8 \\
3 -- 4 & 28 & 34 & 1.21 $\pm$0.3 \\
4 -- 5 & 22 & 28 & 1.27 $\pm$0.35\\
5 -- 6 & 16 & 22 & 1.37 $\pm$0.45\\
6 -- 7 &18  & 24 &  1.33 $\pm$0.4\\
7 -- 8 & 25  & 24 &  0.96 $\pm$0.3\\
2 -- 5 & 60  & 70 &  1.17 $\pm$0.2\\
5 -- 8 & 59  & 70 &  1.18 $\pm$0.2\\
\hline
\end{tabular}
\label{tab:completeness}
\end{table}

\citet[][Section 4]{Karachentsev2004} discuss completeness of the
earlier sample and conclude that within 8~Mpc radius the sample was
70-80 percent complete: about 100 galaxies were estimated to be missed
in that sample. We use the updated sample, which has $\sim 100$ more
galaxies and, thus, it is expected to be nearly complete. We estimate
the completeness of our updated sample using two methods. In both
methods we use the ratio of the number of dwarf galaxies to the number
of bright galaxies as an indicator of completeness: the ratio should
not depend on the distance.
First,  we count the number of bright galaxies
($M_B<-15$) and the number of dwarf galaxies ($M_B=-12-14.5$) inside
radial shells of 1~Mpc width. If the sample is not complete, we would
expect a decline with the distance of the number of dwarf
galaxies. The ratio of the number of dwarf to large galaxies presented
in the Table~\ref{tab:completeness} do not indicate any decline and
confirm the completeness of the sample. Second, we  make the counts of
galaxies in the zone of avoidance and compare them with the counts in
the direction of the galactic pole. For the same two subsamples
($M_B<-15$ and $M_B=-12-14.5$) we find 28 bright galaxies and 18
dwarfs close to the galactic plane ($|b|<15^o$). In the direction of
the galactic pole ($|b|>75^o$) we find 16 dwarfs and 28 giants. This
gives the ratio of dwarfs/bright galaxies equal to 0.64 in the the direction of
the galactic pole and 0.57 in the galactic plane. Again, results are
compatible with the completeness of the sample.

The numbers indicate that there is no large incompleteness.  However,
we cannot exclude a very likely possibility that the sample still
misses few galaxies. This is why we do not use statistics (e.g., the
size of the largest void), which are sensitive to small effects such
as missed few galaxies or few galaxies with wrong distances. Results
presented in section~\ref{sec:cvf} demonstrate stability of our
statistics to variations in the sample and to errors in distances.

\begin{figure}
 \begin{center}
  \includegraphics[width=8.3cm]{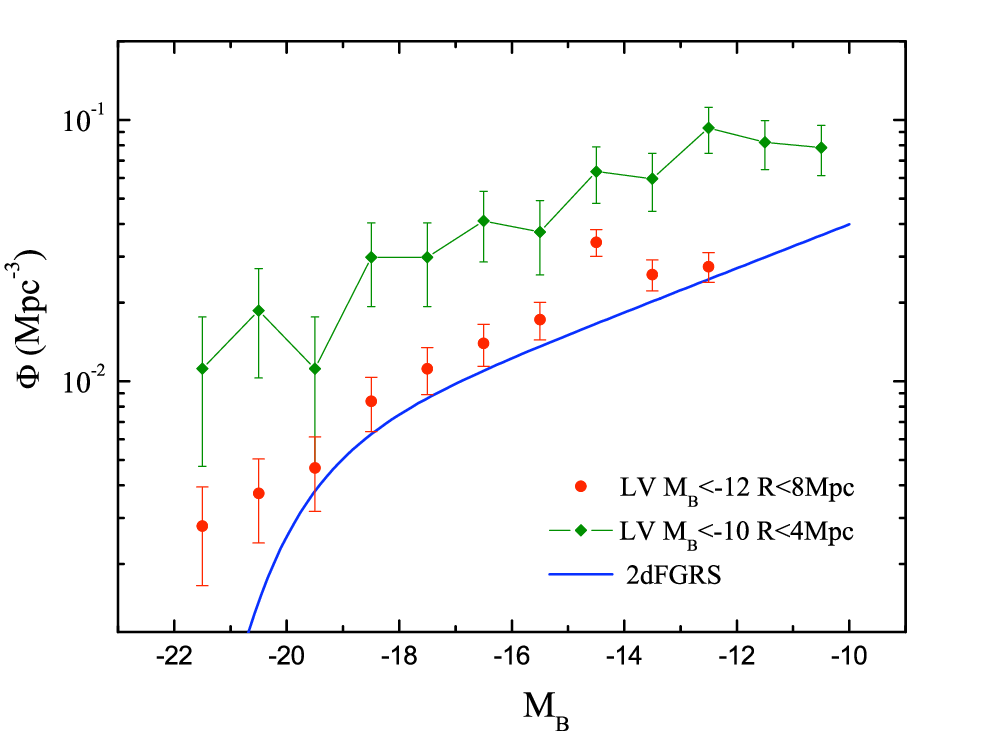}
 \end{center}
 \caption{The luminosity function of galaxies in the the Local
   Volume. Circles with errors show results for the 8\,Mpc sample
   (complete to $M_B=-12$).  The full curve is for the 4\,Mpc sample
   (complete to $M_B=-10$). The dashed curve is for the Schechter
   approximation of 2dFGRS scaled to $h=0.72$. The faint-end slope of
   the approximation $\alpha =1.2$ fits well the slope of the
   luminosity function of galaxies in the Local Volume. For
   illustrative purposes we plot the approximation well outside of
   the $M_B=-17.2$ completeness limit of 2dFGRS.}
\label{fig:LF} \end{figure}

Galaxies in  the Local Volume were detected  down to
extremely low luminosities. This gives us a unique chance to detect
voids, which may be empty of any galaxies.  We use two
volume-limited samples. The main sample is complete for galaxies with
absolute magnitudes $M_B<-12$ within 8\,Mpc radius.  Another volume
limited sample is $M_B<-10$ within 4\,Mpc.

\section{Simulations}

\begin{table*}
\caption{The main parameters of the cosmological simulations}
\begin{tabular}{l|cccc}
\hline
 Simulation       & \BoxS & \BoxCR & \BoxCRgad  \\
\hline
Box Size ($h^{-1}$Mpc) & 80 & 160 & 64 \\
$\sigma_8$ & 0.90 & 0.75 & 0.75 \\
Mass of a high resolution particle ($h^{-1}M_{\sun}$) & $4.91\times 10^6$ & $3.18\times 10^8$ & $1.6\times 10^7$  \\
Spatial Resolution ($h^{-1}$kpc) & 0.52 & 1.2 & 1.6 \\
Number of high resolution particles & $1.6\times 10^8$ & $1024^3$ & $1024^3$  \\
Circular velocity of the smallest resolved halo (km/s)&  9   &  27   &  15   \\
\hline
\end{tabular}
\label{tab:boxes}
\end{table*}

We use three N-body simulations: two are done with the Adaptive
Refinement Tree code \citet{Kravtsov1997}, and one simulation, which
is done with the GADGET2 code \citet{Gadget2}.  Parameters of
simulations are presented in the Table~\ref{tab:boxes}.  The simulation
\BoxS~ has a high-resolution spherical region of radius $10h^{-1}{\rm Mpc}=14$\,Mpc, which is 
resolved with $\sim $150 million particles each having mass $5\times
10^6h^{-1}M_{\odot}$. This high-resolution region is embedded into box
of $80 h^{-1}$~Mpc were mass and force resolutions are lower. In the other two
simulations the whole computational box was resolved with equal-mass
$1024^3$ particles particles.

We use halo finders, which detect  both distinct haloes and their
subhaloes. The halo finders provide us with different parameters of (sub)haloes.
As a measure of how large is a halo we typically use
the maximum circular velocity $V_c$, which is easier to relate to
observations as compared with the virial mass.  For reference, haloes
with $V_c =50$\,km/s have virial mass about $10^{10}M_{\odot}$ and
haloes with $V_c =20$\,km/s have virial mass about $10^{9}M_{\odot}$.
 
The simulations are for a spatially flat cosmological LCDM model with
following parameters. The simulation \BoxS~ has $\Omega_0 = 0.3,
\Omega_{\Lambda} = 0.7; \sigma_8 = 0.9; h = 0.7$ (WMAP1
parameters). Simulations \BoxCR~ and \BoxCRgad~ are done for $\Omega_0 =
0.24, \Omega_{\Lambda} = 0.76; \sigma_8 = 0.75; h = 0.73$ (WMAP3
parameters).

We re-scaled all data (coordinates and masses of haloes) to ``real''
units assuming $H_0 = 72$ km/s/Mpc, which is close to recent
WMAP results.

\section{Results}\label{sec:results}

\subsection{Luminosity function and global parameters of LV}\label{sec:lf}

Figure~\ref{fig:LF} shows the luminosity function of LV galaxies.The
Figure also shows the Schechter approximation for the 2dFGRS catalog
\citep{Norberg02} with parameters $\alpha=-1.21$,
$M_{bj}^*=-19.66+5\log h$, $\Phi_*=1.61\times 10^{-2}h^3$\,Mpc$^{-3}$
scaled for $h=0.72$. The approximation gives the average luminosity
function of galaxies in the $b_j$-band in the Universe scaled to the
B-band by means $b_j = B-0.28(B-V)$ \citep{Norberg02} and by assuming
the mean B-V=0.5 color for giant spiral galaxies dominating the Local
Volume. Note that the 2dFGRS luminosity function extends only down to
$M_B\approx -17.2$. We plot it well below its completeness limit
just for reference.  We construct the LF of the LV using magnitudes
corrected for the internal extinction.  The LF for uncorrected
magnitudes from tables of \citet{Karachentsev2004} doesn't change the
LF for most of the luminosity bins except for the very brightest one.

The luminosity function of the 8\,Mpc sample is larger than the
universal LF for all luminosities. With the exception of the very
brightest galaxies, the LF in the Local Volume exceeds the universal
LF by a factor of 1.3. The LF of the 4\,Mpc sample gives significantly
larger excess over the universal LF: factor 3.6 for $M_B>-19$. Another
interesting feature of LV $M_B<-12$ luminosity function is the
prominent peak at $M_B=-14$. It is mostly produced by dwarf irregular
isolated galaxies, which have large gas masses, specific SFRs, and total
mass-to-light ratios.  The overall shape of the LF in the Local Volume
is surprisingly robust. For example, the excess of the bright galaxies
with $M_B< -20$ is reproduced in both subsamples. The slope of the LF
at the faint end $\alpha \approx -1.2$ is also the same for the two samples.

In addition to the overall normalization of the LF, we also estimate
the overdensity for the LV sample using a slightly different method.
We find the total luminosity of galaxies in LV in the range
$-22<M_B<-17$ and compare it with the expectations from the 2dFGRS. In
this case we average uncertainties due to the small number statistics
at the very brightest end of the LF and use the 2dFGRS luminosity
function within its limits of completeness.  We obtain the following
overdensity ratios: in the 8\,Mpc sample the overdensity is $1.4 \pm
0.17$; in the 4\,Mpc sample the overdensity is $\sim 5.3$.  

Note that the spherical top-hat collapse model predicts that a sphere
with overdensity ~5.5 is at the turn-around radius \citep{Peebles1980}
and does not expand.  If the model were applicable to the motion of
galaxies in the Local Volume, than
we should have observed random motions comparable to the Hubble
velocities for the 4\,Mpc sample. In reality we have rather cold Hubble flow
in the Local Volume implying that the top-hat model gives an extremely
poor approximation of the dynamics of galaxies in the LV.

Galaxy overdensities, which we find in the LV, give us additional
criteria for selection of LV-candidates in simulations.  Since a large
fraction of the number and luminosity overdensities in the LV are due
to bright galaxies, we use the observed number overdensities as a
condition for the overdensity of large haloes (circular velocity $V_c >
100km/s$) in 8\,Mpc sphere. There is another constraint, which limits
the number of galaxies at small distances. There are no bright
galaxies just outside of  the Local Group and up to the distance of
3\,Mpc. 

\subsection{Selecting model LV-candidates}
To mimic the main features of the real Local Volume galaxy sample we use a
number of criteria to select from our simulations spheres of radius
8\,Mpc that we can consider as ``LV-candidates''. Criteria differ
slightly from one simulation box to another to select at least some
candidates. In all simulations the candidates must be centered on a
halo with the virial mass in the range $1.5\times10^{12}M_{\odot} <
Mass < 3\times 10^{12}M_{\odot}$.  We do not require that the
LG-candidates should have two comparable in mass haloes. Appearance of
two haloes instead of one halo of the same total mass does not change
the dynamics of matter outside the LG-candidate. We also do not
require that the candidates should be at the same distance from a
Virgo-type cluster as the real LG. About a half of our LG-candidates
are at close ($15-25$Mpc) distance from a cluster. We do not find any
substantial difference between those candidates. Instead, we impose
detailed constraints on distribution of mass in the LG-candidates,
which we take from the real LV sample of galaxies.  Below is the list
of conditions, which were used for different simulations:

{\it Simulation \BoxS: } (1) no haloes with mass$> 4\times
10^{13}M_{\odot}$ inside a 8\,Mpc sphere. Thus, no large groups and
clusters in a sample.; (2) The number density of haloes found inside
8\~Mpc sphere with $V_c>100$\,km/s exceeds the mean value in the whole box
by factor in the range $1.3-1.7$; (3) There are no haloes more massive
than $8\times10^{11}M_{\odot}$ with distances in the range (1-3\,Mpc).
In total, there are 3 LG-candidates in the simulation \BoxS, which satisfy
these conditions.

{\it Simulation \BoxCRgad:} (1) no haloes with Mass$> 2\times
10^{13}M_{\odot}$ inside a 8\,Mpc sphere; (2) The number density of
haloes found inside 8\~Mpc sphere with $V_c>100$\,km/s exceeds the mean
value in the whole box by factor in the range $1.5-1.7$; (3) There are
no haloes more massive than $1.0\times 10^{12}M_{\odot}$ with distances
in the range 1-3\,Mpc. (4) Central haloes of different LG-candidates
are located at distance more then 5\~Mpc one from the other. There are
14 samples with above criteria in the simulation \BoxCRgad.

{\it Simulation \BoxCR:} The volume of the simulation is large enough
to allow us to select candidates, that mimic LV features more closely
than in the previous simulations. Conditions are: (1) no haloes with
$Mass > 2\times10^{13}M_{\odot}$ inside a 8\,Mpc sphere; (2) The
number density of haloes found inside 8\~Mpc sphere with
$V_c>100$\,km/s exceeds mean value in the whole box by factor in the
range $1.5-1.7$; (3) The number density of haloes found inside 4.5\~Mpc
sphere with $V_c>100$\,km/s exceeds the mean value in the whole box by
factor greater then 3.5; (4) There are no haloes more massive than
$5.0\times 10^{11}M_{\odot}$ with distances in the range
(1-3\,Mpc). 
(5) Central haloes of different LG-candidates are
separated by distances larger than 8\~Mpc . We've found 7 such
candidates in the simulation \BoxCR.

We want to note that our LV-candidates look quite similar to the LV
galaxy sample.  They have the same environment, and they have the same
mixture of groups (or lack of those).  The number of large galaxies is
also very similar. In the observed sample there are about 20 galaxies
with rotational velocities larger than 120~km/s.  On average, there
are about 20 haloes with circular velocity above 100~km/s in our
LV-candidates in the simulations with the WMAP3 parameters.  Taking
into account the increase in rotational speed produced by baryonic
infall into DM haloes, we expect the models produce the same number of large ``galaxies''
as observed in the real LV. For all selected LV-candidates we estimate the
rms velocity deviations from the Hubble flow and find that they are
reasonably consistent with values obtained from LV galaxy sample.

\begin{table*}
 \centering
   \caption{Velocity scatter around the Hubble flow in Local Volume as
     the function of the distance from the Local Group.}
 \begin{tabular}{c|ccccccc}
  \hline
Outer bin     & Number of & Uncorrected  & Corrected for& Corrected for&Corrected for &Apex & error\\
 radius         & galaxies    &  rms velocity& distance errors & apex motion&all effects &velocity & \\
$D_{\rm out}$ (Mpc)  & N  &  $\sigma_H^0$ & $\sigma_H^1$  &   $\sigma_H^{a}$  &  $\sigma_H^{f}$ & $V_{apex}$(km/s) & $\sigma_{m}$\\
\hline

3.0& 43  &  73.7  &    71.7  &    56.2 &  53.5 & 65.5 & 17.4\\
4.0& 106 &  84.6  &     81.6 &    83.4 &  80.3 & 26.7 & 22.4\\
5.0& 162 &  84.3  &     80.1 &    83.1 &  78.8 & 21.9 & 26.3\\
6.0& 214 &  83.6  &     78.1 &    81.2 &  75.4 & 32.0 & 29.9\\
7.0& 273 &  96.8  &     90.4 &    90.6 &  83.8 & 54.9 & 34.4\\
8.0& 335 &  106.6 &     99.3 &    98.2 &  90.2 & 68.8 & 38.8\\
9.0& 360 &  107.5 &     99.6 &    99.4 &  90.7 & 68.5 & 40.7\\
\hline
\end{tabular}
\end{table*}

\subsection{Rms peculiar velocities and the ``coldness'' of the Hubble flow}
One of the major challenges for the cosmological models was to
reproduce the value of peculiar velocity dispersion $\sigma_H$ in
Local Volume, where the Hubble flow appeared to be rather
``cold''. For the estimation of $\sigma_H$ in galaxy samples we use
computational scheme that differs from that used by Karachentsev's
group. In order to have a simple and clear interpretation of obtained
values and direct and unbiased comparison with simulated
LV-candidates, we take all galaxies with known radial velocities in
the frame of Local Group (placing the centre of our samples just in
the middle between Milky Way and M31). We assume that the Hubble flow
is universal and equal to the universal cosmic expansion. We do not
correct results for virial motions due to groups inside the LV: we do
not exclude any galaxies -- such as members of groups -- from
statistics. Removing these velocities from the estimates of peculiar
velocities and finding the Hubble flow from the observed galaxies in
the LV underestimates the value of $\sigma_H$. 

 We assume
a 10\% error as the mean error of distance measurements. We start our
estimation just outside 1\,Mpc radius from the centre of the Local
Group and calculate $\sigma_H$ for galaxies with distances $D$ from
$1Mpc$ to the distance $D_{\rm out}$. We apply a correction for the
apex motion -- the mean motion of galaxies in surrounding volume with
respect to the Local Group in simple dipole approximation (making our
$\sigma_H$ estimation in the frame where surrounding galaxies have no
dipole component motion relative to LG). The vector of apex motion
\{$ A_x, A_y, A_z$\} is estimated by minimization of the sum:
\begin{equation}
\sum_{i=1}^N \left(v_i - H_0 D_i - \frac{A_x x_i + A_y y_i +
A_z  z_i}{D_i}\right)\,,
\end{equation}
where $H_0 = 72$ km/s/Mpc and $v_i$, $D_i$, $x_i$, $y_i$, $z_i$ are
velocity, distance and Cartesian coordinates of galaxies in the Local
Group frame. After subtracting the apex and the Hubble flow ($H_0
D_i$) from velocities, we calculate the residual rms peculiar radial
velocity in a certain distance range. By subtracting apex from
velocities (converted into LG frame) we remove the dipole component of
motions of galaxies.  Now we need to correct rms peculiar velocities
for the errors in distances of galaxies. We subtract in quadratures
the estimated rms velocity $\sigma_{\rm m}$ due to the errors:

\begin{eqnarray}
& \nonumber \disp \sigma^2_{\rm m} = \frac{1}{N}\disp
\sum_{i=1}^N(\Delta v_i + H_0 \delta D_i)^2 = \frac{1}{N}\disp
\sum_{i=1}^N\Delta v_i^2
+ &\\
& + \frac{\alpha^2 H^2_0}{N} \disp \sum_{i=1}^N D^2_i + \disp
\frac{2 H_0}{N}\disp \sum_{i=1}^N ( \Delta v_i \delta D_i),&\,
\end{eqnarray}
where we assume that the error in the peculiar radial velocity
$\sigma_{m}$ is produced by errors in measurements of velocities
$\Delta v_i$ and distances $\delta D_i$. Here the parameter $\alpha$
is the rms error in distance measurements: $\delta D_i = \alpha D_i$,
$\alpha \approx 0.1$.  We assume that the third term in the last
equation is small: no correlation of errors of peculiar velocities and
distance errors. Our Monte-Carlo modeling of the errors shows that
this is the case.  We also neglect the first term in the right-hand
side of the equation (2) since velocity measurements are accurate
enough ($\Delta v_i \approx 5$\,km/s).

\begin{figure}
 \begin{center}
  \includegraphics[width=8.3cm]{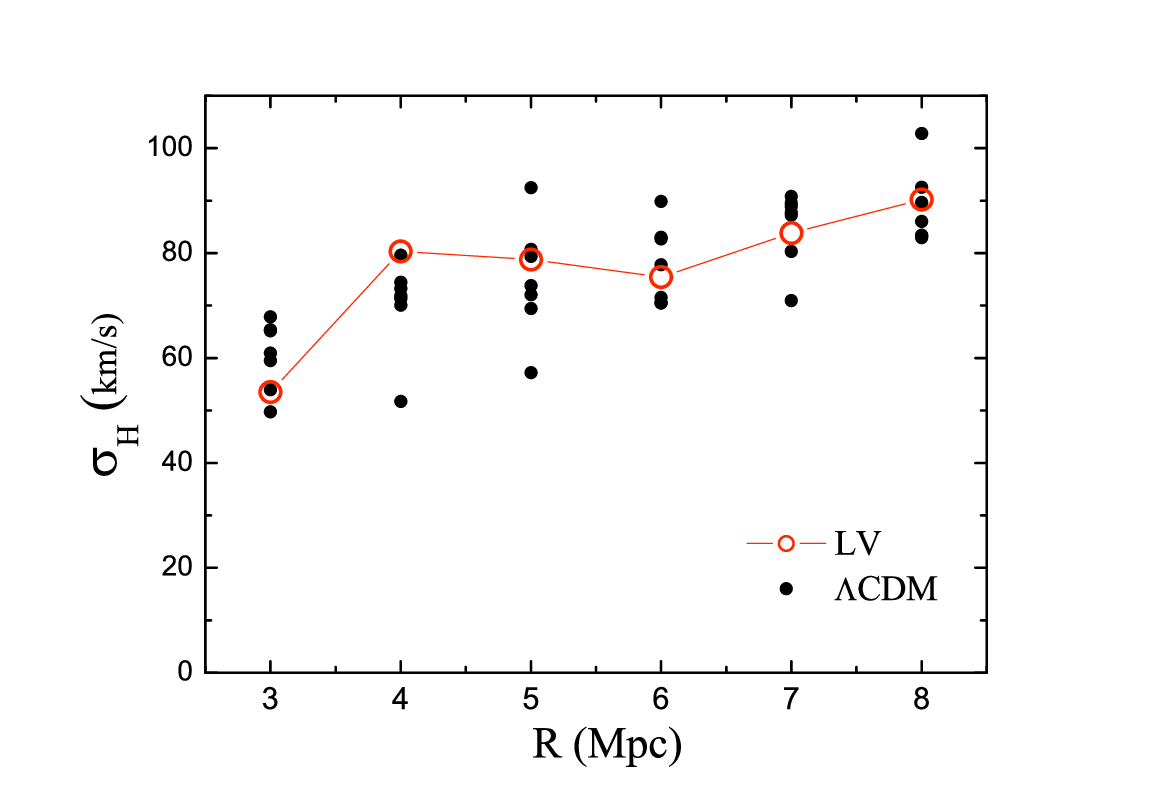}
   \end{center}
   \caption{The rms radial velocity deviations from the Hubble flow
     $\sigma_H$ for galaxies in the Local Volume with distances from
     1\,Mpc up to R (full red curve with open circles). The estimates
     are corrected for the apex motion and for distance errors. Black filled
     circles show theoretical predictions for 7 LV candidates in the
     simulation \BoxCR. }
\label{fig:velrms} \end{figure}

Values of $\sigma_H$ on different scales for galaxies in the Local
Volume and complementary parameters are presented in Table~2.  The
table gives the rms velocity in different overlapping regions of LV
for distances from 1\,Mpc to $D_{\rm out}$. The second column gives
the number of galaxies $N$ in a region limited by $D_{\rm out}$ and
with distance larger than 1\,Mpc.  The parameter $\sigma_H^0$ is the
radial rms deviation from Hubble flow with no corrections for either
apex or distance errors. The parameter $\sigma_H^1$ is the rms
deviations corrected for distances errors: $\sigma_{m}$ (last column)
is estimated using the equation (2) and subtracted in quadratures from
$\sigma_H^0$.  Column 5 gives the rms velocity corrected for apex
only.  Parameter $\sigma_H^{\rm final}$ (column 6) is the final
estimate of the rms radial deviations: corrected for apex motion and
for distance errors.  Two last columns give the velocity of apex
motion and the error estimated by eq.(2). Figure~\ref{fig:velrms} shows
$\sigma_H^{\rm final}$ in graphical form (the red curve with circles).

When analyzing LV-candidates in the simulations, we subtract 3d
velocity of the central halo from the rest of sample velocities, find
radial velocities of haloes and apply the above procedure of apex
correction and Hubble flow subtraction and then obtain
$\sigma_H^{model}$. In this case the apex motion is just the dipole
component of motions of haloes with respect to the central halo --
LG-analog. Results for the simulation \BoxCR~ are shown in
Figure~\ref{fig:velrms}~ with black circles. Results of the simulation
\BoxCRgad~ are very similar.

As can be seen from Figure~\ref{fig:velrms}, the theoretical
predictions are close to the observed values.  We note the importance
of criterion that between 1\, Mpc and 3\,Mpc there are no relatively
large DM haloes. With such a condition  we get
the mean rms velocity $\sigma_H^{model}\sim
50$\,km/s for 8\,Mpc samples  and for the distances less than 3\,Mpc in the simulation
\BoxCR. This is very close to the observational value.  For some
samples (not shown on the plot) $\sigma_H^{model}$ is as low as about
20\,km/s.  If we don't apply this condition and allow large haloes with mass
$\sim 10^{12}M_\odot$ to reside between 1 and 3\,Mpc, then the mean
$\sigma_H^{model}$ in this region is as high as 72.3\,km/s. So
 massive haloes heat the ``halo gas'' leading to large rms velocities.

\begin{figure}
 \begin{center}
  \includegraphics[width=8.3cm]{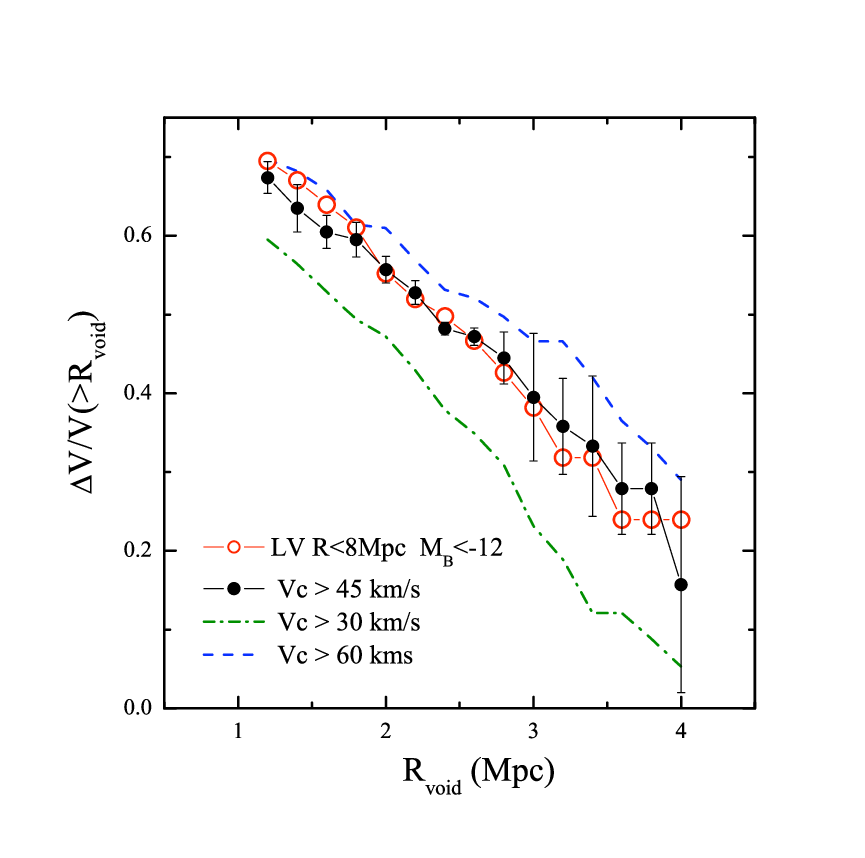}
   \end{center}
   \caption{Fraction of volume $\Delta V/V$ in voids with radius
     larger than $R_{\rm void}$.  Observational data (the complete
     sample, open circles) are compared with the distribution of voids
     in samples of haloes with different limits on halo circular
     velocity in the simulation \BoxS~ with $\sigma_8=0.9$. VF for $V_c=45$\,km/s provides
     a remarkably good fit to observations. Note that the Local Volume
     has very large empty regions with about 1/4 of the whole
     volume occupied by one  void.}
\label{fig:vf80s} \end{figure}

\subsection{Detecting Voids}\label{sec:voids}
In order to detect voids, we place a 3d mesh on the observational
or simulation volume. We then find initial centres of voids as the
mesh centres having the largest distances to nearest objects. In
the next iteration, an initial spherical void may be increased by
adding additional off-centre empty spheres with smaller radius.
The radius of the spheres is limited to be larger then 0.9 of the
initial sphere radius $R_{seed}$ and their centres must stay
inside the volume already connected to void. The process is
repeated until $R_{seed}=1$\,Mpc. It produces voids which are slightly
aspherical. Mean ellipticity of our voids is about 0.7. Artificial
objects are placed on the boundaries of the sample to prevent
voids getting out of the boundaries of the sample.  We define the
cumulative void function (VF) as the fraction of the total volume
occupied by voids with effective radius larger than $R_{\rm eff} =
(3 V_{\rm void}/4\pi)^{-1/3}$ (further $R_{\rm void})$. As we already
mentioned, for our purpose -- to match observational VF by model
one with a certain limit on circular velocity -- the voids volume
statistic (VF) is a robust one since it is not very sensitive to the
total number of objects in a sample.

\subsection{Void Functions}\label{sec:cvf}

\begin{figure}
 \begin{center}
  \includegraphics[width=8.3cm]{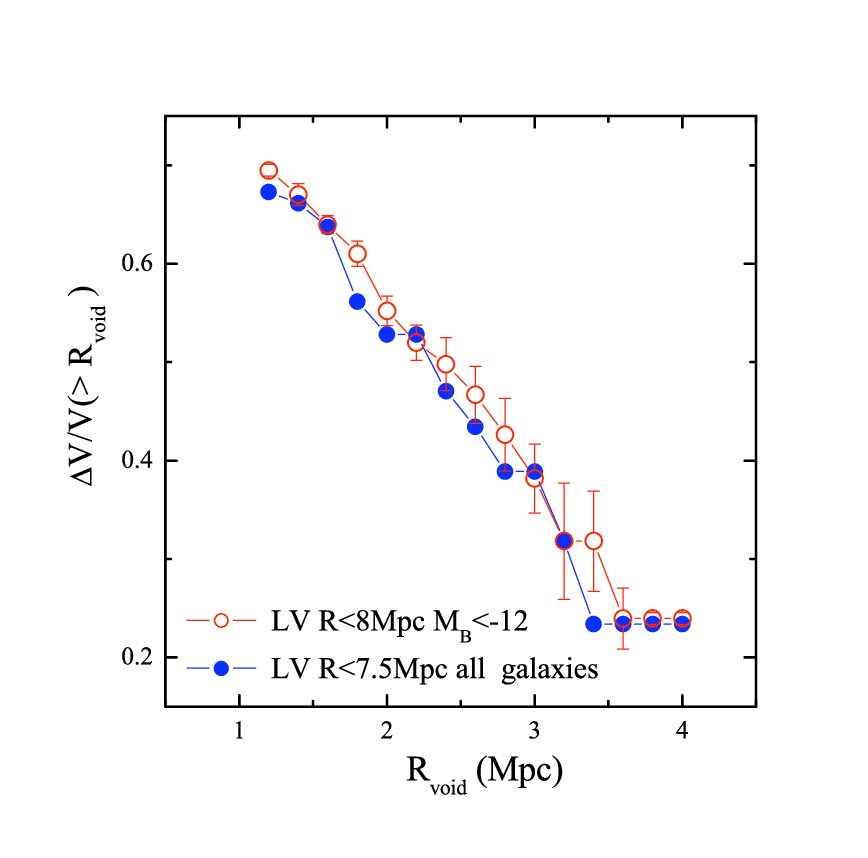}
   \end{center}
   \caption{The void function for two observational samples. The full
     curve with open circles are for a complete volume limited sample
     with $M_B<-12$ and $R<8$\,Mpc. $1 \sigma$ errors obtained by
     Monte Carlo re-sampling of distances from catalog by means of
     addition of Gaussian radial displacements with typical distance
     error of 10\% of distance measurements.  The filled circles are
     for all observed galaxies inside 7.5\,Mpc.  Comparison of the
     samples shows reasonable stability of the void function.}
\label{fig:lvvf} \end{figure}

\begin{figure}
 \begin{center}
  \includegraphics[width=8.3cm]{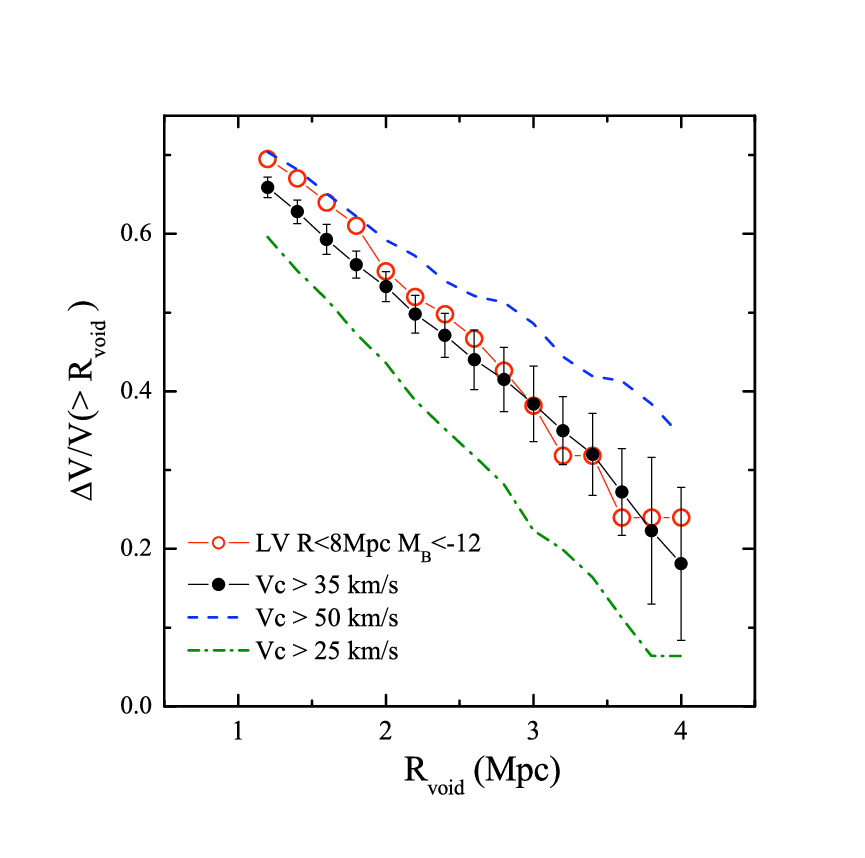}
   \end{center}
   \caption{The observational void function (the complete sample with
     $M_B<-12$) is compared with the distribution of voids in 14
     samples in the simulation \BoxCRgad~  for haloes with different limits on halo
     circular velocity. In this case VF for $V_c=35$\,km/s (shown with
     $1 \sigma$ scatter) provides a better fit to observations.}
\label{fig:vf64} \end{figure}

\begin{figure}
 \begin{center}
  \includegraphics[width=8.3cm]{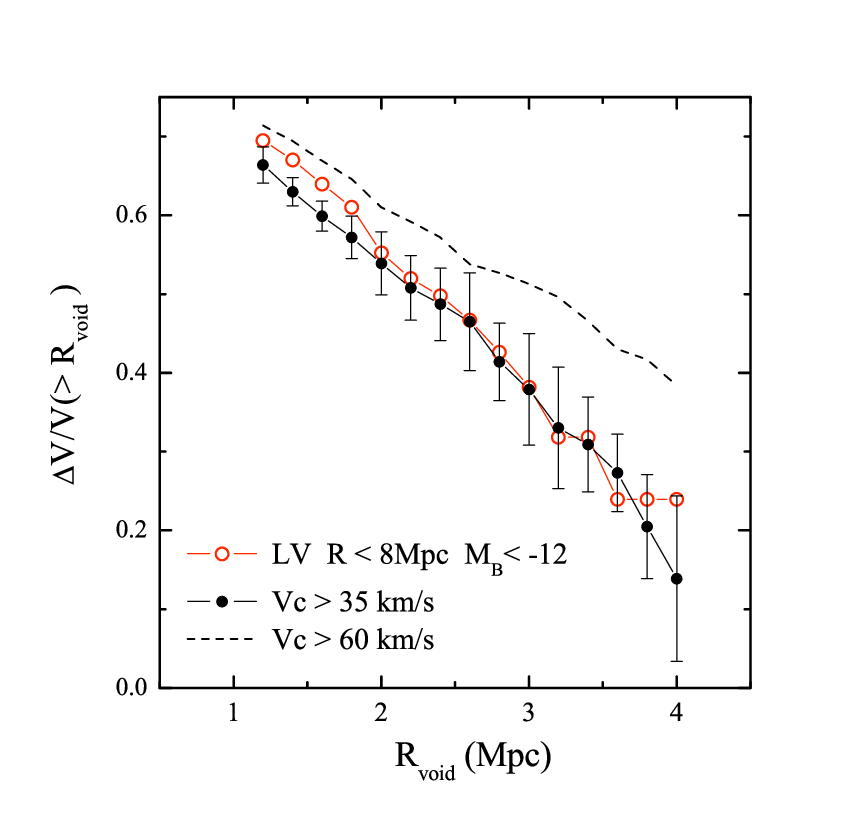}
   \end{center}
   \caption{Observational data (the complete sample $M_B<-12$) are
     compared with the distribution of voids in 7 samples from \BoxCR~
      simulation for haloes with different limits on circular velocity. VF for
     $V_c=35$\,km/s (shown with $1 \sigma$ scatter) provides a
     remarkably good fit to observations. Because of resolution
     limitations in this simulation, we do not present results for
     smaller $V_c$ limits.}
\label{fig:vf160} \end{figure}

We use two samples to construct VF of the Local Volume: (1) Galaxies
brighter than $M_B=-12$ inside sphere of radius 8\,Mpc; the number of
galaxies is 315.  and (2) all galaxies inside 7.5\,Mpc; the number of
galaxies is 376. Results are present in the
Figure~\ref{fig:lvvf}. There are about 30 voids in the observational
samples with radii $1-4.5$\,Mpc. We limit the radius of voids to be
more than 1\,Mpc. The two subsamples indicate some degree of
stability: inclusion of a number of low-luminosity galaxies does not
change the void function significantly.

The Figure~\ref{fig:vf80s} shows the mean VF for 3 different samples
of haloes from the simulation \BoxS~ and the observed VF. Results indicate that
voids in the distribution of haloes with $V_c
> 45$\,km/s give the best fit to the observed VF: the spectrum of voids in the
most valid range of $R_{void}$ is reproduced by the theory. The
theoretical VF goes above the observational data, if we use
circular velocities larger than 60\,km/s. If we use significantly
lower limits, than the theory predicts too few large voids. The
theoretical results match the observations, if we use $V_{circ} =
45 \pm 5$\,km/s.

The Figure~\ref{fig:vf64} and Figure~\ref{fig:vf160} show the mean VF
for 14 and 7 different samples of haloes from simulations \BoxCRgad~
and \BoxCR~ respectively and the observed VF. Here voids in the
distribution of haloes with $V_c > 35$\,km/s match the observed
VF. The difference in limiting value of $V_c$ between the simulations
is related to different values of $\sigma_8$ used in simulations. The
theoretical results match the observations, if we use $V_c = 35 \pm
5$\,km/s. Note that 7 LV-candidates used here are those whose
$\sigma_H$ values plotted on Figure~\ref{fig:velrms}: these samples
mimic LV environment more closely.

\begin{figure}
 \begin{center}
  \includegraphics[width=8.3cm]{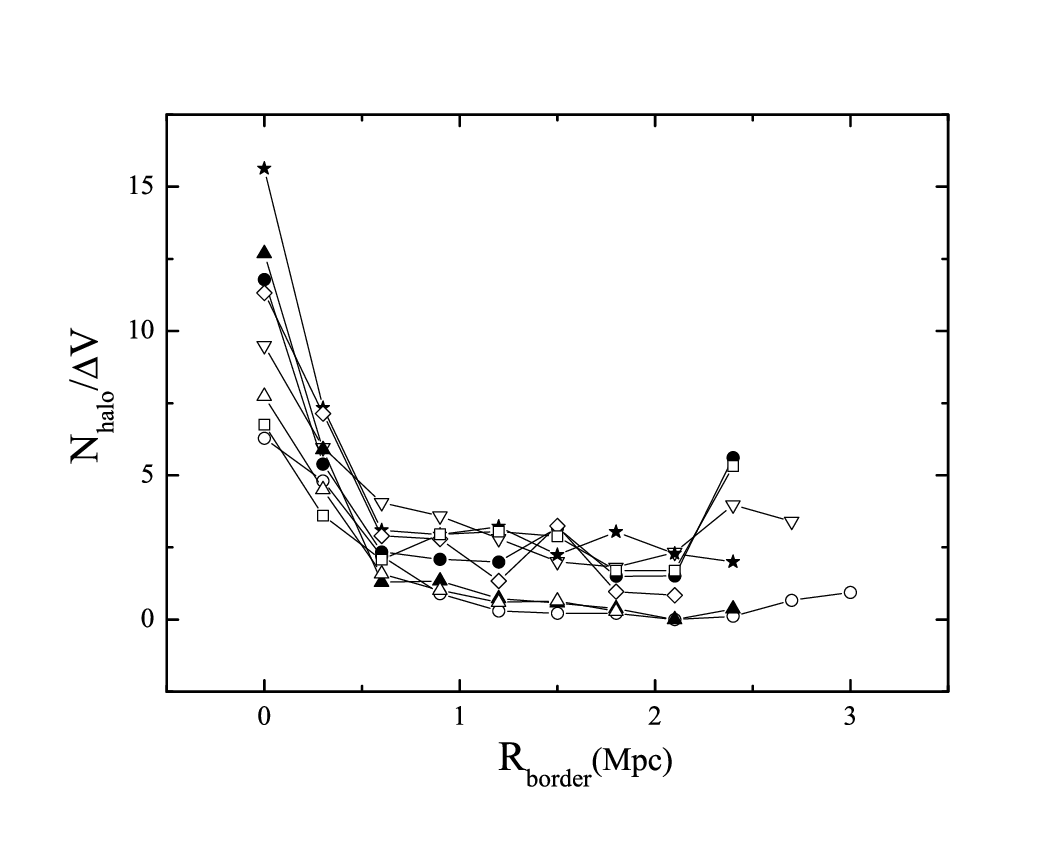}
   \end{center}
   \caption{The number density profiles $N/\Delta V$ of haloes with
     $V_c < 45$ km/s inside 8 largest voids (different symbols) found
      in the \BoxS~ simulation. The distance $R_{border}$ is the
     distance of a halo from the void boundary. The haloes are binned
    in 300~kpc shells.}
\label{fig:prof} \end{figure}

We study the distribution of very small haloes inside voids defined by
larger haloes.  We use 8 largest voids in the simulation \BoxS, which
are defined by haloes with $V_c>45$\,km/s.  There are smaller haloes
inside the voids. We characterize the haloes by their distance $R_{\rm
  border}$ to the border of a void. (Note that $R_{\rm border}=0$ is
for a halo at the boundary of a void, not at its centre).  We count
the number of the haloes in shells with width 300~kpc. The
Figure~\ref{fig:prof} shows the number density profile of haloes as the
function of $R_{\rm border}$. The number density of haloes is very low
close to the centres of voids and increases very substantially when we
get closer to the void boundary. In this respect the small voids,
which we study in this paper, behave very similar to giant voids found
in simulations of \citet{Gottloeber2003}.  To a large degree, the
small and giant voids are similar. For example, there are very small
filaments made of tiny haloes inside our small voids. 
Altogether we have a kind of self-similarity of voids
properties on different levels of their detection both in observations
and in simulations \citep{Patiri2006, Tikh07, Gottloeber2003}.

\section{Discussion and Conclusions}\label{sec:concl}
\begin{table*}
 \centering
   \caption{Properties of isolated dwarf galaxies with $M_B=-11.8-13.2$}
 \begin{tabular}{l|cccccl}
  \hline
Name     & $M_B$ & axial ratio  & $W_{50}$& $V_{\rm rot}$& Distance &reference \\
\hline
E349-031,SDIG  &    -12.10  &   0.82  &   20.0   &    17.5  &      3.21&  \citet{Karachentsev2004}  \\
KKH5         &      -12.27 &    0.62  &   37.0   &    23.6  &      4.26 &  \citet{Karachentsev2004}\\
KKH6          &     -12.38 &    0.60  &   31.0   &    19.4  &      3.73 &  \citet{Karachentsev2004} \\
KK16           &    -12.65  &   0.37  &   24.0   &    12.9  &      5.40  & \citet{Karachentsev2004}\\
KKH18        &      -12.39  &   0.57  &   34.0  &     20.7 &       4.43  & \citet{Karachentsev2004}\\ 
KKH34,Mai13    &    -12.30  &   0.56  &   24.0   &    14.5  &      4.61 &  \citet{Karachentsev2004}\\
E489-56,KK54  &     -13.07  &   0.53  &   33.8  &   19.9  &            4.99  & \citet{HIPASS}\\
KKH46       &       -11.93  &   0.86  &   25.0   &    24.5  &      5.70  & \citet{Karachentsev2004}\\
U5186       &       -12.98  &   0.23  &   42.0  &     21.6  &      6.90  & \citet{Karachentsev2004}\\
E321-014   &        -12.70  &   0.43  &   39.8 &    22.0    &        3.19  & \citet{HIPASS}\\
KK144        &      -12.59 &    0.33  &   44.0  &     23.3  &      6.30  & \citet{Karachentsev2004}\\
E443-09,KK170   &   -12.03  &   0.75 &    29.0  &     21.9 &       5.78  & \citet{Karachentsev2004}\\
KK182,Cen6   &      -11.89  &   0.60  &   16.0   &    10.0&        5.78  & \citet{Karachentsev2004}\\
DDO181,U8651   &    -12.97  &   0.57  &   42  &    23.7  &      3.02  & \citet{Karachentsev2004,Springob05}\\
DDO183,U8760   &    -13.13  &   0.32  &   30.0  &     15.8 &       3.18  & \citet{Karachentsev2004}\\
HIPASS1351-47  &    -11.88  &   0.60  &   38.8  &     24.2   &         5.65 &  \citet{HIPASS} \\
\hline
\end{tabular}
\end{table*}

\begin{figure}
 \begin{center}
  \includegraphics[width=8.3cm]{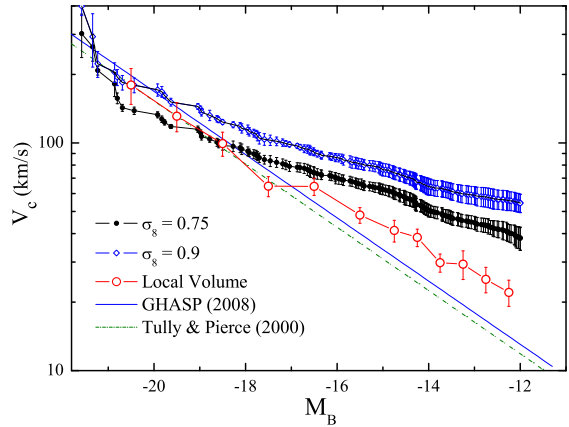}
   \end{center}
   \caption{The velocity-magnitude relation for galaxies in the Local
     Volume  (open circles with error bars) is compared with
     predictions of the LCDM model. Two other observational estimates
     are also shown  as full and dashed lines. The theory (filled circles and diamonds)
     makes reasonable predictions for bright galaxies with
     $M_B<-17$. At smaller luminosities the theoretical curves are
     systematically  above the observations. At $M_B=-12$ the
     disagreement is a factor of two in circular velocities implying a
     factor of $\sim 10$ disagreement in the number of haloes.}
\label{fig:TF} \end{figure}

We use an updated version of the  \citet{Karachentsev2004}  sample of
galaxies to study the distribution and motions of galaxies in the Local Volume.
There are about 30 voids, which range in
radius from 1\,Mpc to 4\,Mpc. We demonstrate that the spectrum of void sizes
is relatively stable for variations of the sample and for uncertainties
of distances to individual galaxies. Estimates of the cosmic variance,
which we get from cosmological simulations, also show stability of the
void statistics.

When making the theoretical predictions for the LCDM model, we
carefully select Local Volume candidates. In many respects the
candidates look very similar to the reality: they have similar number
of large haloes, similar density contrasts at different scales, and
have similar rms velocity deviations from the Hubble flow.  The
spectrum of void sizes in simulations traces the observed spectrum
remarkably well, if we assume that haloes with circular velocities
$V_c>35$~km/s for $\sigma_8=0.75$ and $V_c>45$~km/s for
$\sigma_8=0.90$ host galaxies brighter than $M_B=-12$. The mass limits
are quite consistent with the theoretical expectations for the mass of
smallest halo, which can host a galaxy
\citep[e.g.,][]{Hoeft06,Loeb08}.

At the same time, if much smaller haloes with $V_c>20$~km/s host
galaxies with the observed absolute magnitudes $M_B=-12$, voids in the
LCDM model would be too small and their spectrum of sizes would
strictly contradict the observations.  This is hardly an unexpected
conclusion: in the hierarchical scenario any void is filled with small
haloes. The only question is what is the mass of the haloes. For the
Local Volume with the completeness limit $M_B=-12$ this appears to be
$V_c>35-45$~km/s.  If this is true, haloes with $V_c\approx 20$\,km/s
should not host galaxies. {\it The problem is that in reality they do}: in
the Local Volume many luminous galaxies with these absolute
magnitudes  rotate with velocities $V_{\rm rot}\approx 20$\,km/s
or smaller.

In order to demonstrate this, we select all isolated galaxies with
$-13.2<M_B<-11.8$. The isolation criteria are very strict because we
would like to be sure that the rotational velocities of galaxies were
not reduced by stripping or by any other interaction with large
neighboring galaxies. We select the galaxies, which are more than
1\,Mpc away from any galaxy brighter than $M_B<-19$ and do not have
companions within 200~kpc, which are brighter than the galaxy
itself. Note that the galaxies are so small that the expected virial
radius is smaller than 100\,kpc. All the galaxies are dwarf
irregulars, and for half of them there are measurements of the 21\,cm
HI line width.  Using the HI full-width-half-maximum $W_{50}$
measurements, we estimate the rotational velocity of a galaxy: $V_{\rm
  rot} = W_{50}/2\sqrt{1-(b/a)^2}$, were $b/a$ is the axial ratio. A
large fraction of the HI line width is likely produced by random
8--10~km/s velocities. We do not subtract those  because
we compare the results with the circular velocities of dark matter
haloes. Table~4 presents the results for galaxies with detected HI
emission and with $V_{\rm rot} < 25$~km/s.  Galaxies in the Table~4
have rotational velocities well below those required by the LCDM
model.  Half of the galaxies rotate slower than 20~km/s. We also
studied galaxies, which are not so isolated and galaxies, which are
slightly brighter than those presented in the Table~4. Results are
qualitatively the same.

The disagreement between the theory and observations also shows up in
the Tully-Fisher (TF) relation for galaxies in the Local Volume. In
order to construct the relation, we use all galaxies in the
\citet{Karachentsev2004} sample with measured HI line width $W_{50}$,
which have morphological type $ > 0$ (i.e. spirals and irregulars). The
rotational velocities are corrected for inclinations. In order to
allow the comparison with  theory, we do not make any corrections
for internal gas motions. Open circles with error bars in
Figure~\ref{fig:TF} show the TF relation of galaxies in the Local
Volume. Two lines in the plot also show the TF relation from
\citet{TF2000} and \citet{GHASP}, which we extrapolate down to small
magnitudes (GHASP results extend down to $M_B\approx -16$). The
observational TF estimates agree reasonably well for bright galaxies
with $M_B<-16$. At smaller luminosities the LV results go slightly
above the extrapolations from brighter samples. 

When assigning luminosities to dark matter haloes we follow the
prescription of \citet{Conroy2006}. Specifically, we rank by
luminosity all the galaxies in the LV sample and we rank by circular
velocity all the haloes in our LV-candidate samples in simulations.  We
then take the luminosity of the brightest galaxy and assign it to the
halo with the largest circular velocity. Then we take the second brightest galaxy and
give its luminosity to the second halo and so on. According to
\citet{Conroy2006}, this prescription reproduces clustering properties
of galaxies in the SDSS sample. We add a small ($20\%$) correction to
the circular velocity of haloes to accommodate the effect of adiabatic
contraction due to infall of baryons. Figure~\ref{fig:TF} shows that
the LCDM model gives a good match to observations at the bright end of
the luminosity function ($M_B<-17$). The model with lower
normalization produces a better fit, but even the high normalization
model cannot be excluded: a more accurate treatment of the adiabatic
infall may slightly improve the situation. At low luminosities the
theory and observations gradually diverge, and at $M_B=-12$ the
differences are quite substantial: a factor of two in circular
velocities. This is the same problem, which we found using the
spectrum of voids: haloes with $V_c\approx 35-45$~km/s should have
luminosities $M_B=-12$ in order to match the observational data.

We would like to emphasize that {\it the disagreement with the theory is
staggering.} The observed spectrum of void sizes disagrees at many
sigma level from the theoretical void spectrum if haloes with
$V_c>20$\,km/s host galaxies brighter than $M_B=-12$. We can look at
the situation from a different angle. In the LCDM model with
$\sigma_8=0.9$ there are $\sim 320$ haloes with $V_c>45$\,km/s -- the
same number as the number of galaxies in the Local Volume with the
$M_B=-12$ limit. In the same volume in the LCDM model there are $\sim
3500$ haloes with $V_c>20$\,km/s. If all these haloes host galaxies
brighter than $M_B=-12$, the theory predicts a factor of ten
more haloes as compared with the observations.

The problem has the same roots as the overabundance of substructure in
the Local Group: the LCDM model predicts too many dwarf dark matter
(sub)haloes as compared with the observed dwarf galaxies
\citep{Klypin99, Moore99, Madau08}. We suggest that the solution of
the problem of the overabundance of the dwarfs in the Local Volume
may be similar to current explanations of the substructure problem in the LG:

\begin{itemize}
\item{}The observational sample is not complete: there are ten times
  more dwarf galaxies down to limiting magnitude $M_B=-12$ than listed
  in the \citet{Karachentsev2004} sample. The ``missed'' dwarfs are
  unlikely to be dwarf irregulars because they would have HI emission
  and would have been detected by blind HI surveys such as HIPASS
  \citep{HIPASS}. Dwarf spheroidal galaxies is a possibility. They do
  not have gas and cannot be detected in HI. They have very low
  surface brightness, which makes it difficult to detect them on
  photographic plates. So, it is likely that many of the galaxies were
  missed. Still, we do not know whether a large population of dSph
  galaxies exists in the LV.  If this is so, we will have another
  problem: how to form thousands of dwarf spheroidals in very low
  density environments without any tidal stripping or interaction with
  massive parent galaxy. The slope of the luminosity function also
  will be much steeper: $\alpha \approx 2-2.5$.
\item{} The observed galaxies with $V_{\rm rot}\approx
  20$\,km/s are hosted by significantly more massive haloes. The
  overabundance problem would be solved, if the circular velocity of a
  dark matter halo is $V_c \approx 2V_{\rm rot}$. This is somewhat
  similar to the solution of the overabundance problem in the LG
  \citep[e.g.,][]{Penarubbia08}
\item{} Most of the dwarf haloes with $V_c<35$\,km/s in local voids failed to form stars
  because they collapsed after the epoch of reionization \citep{Bullock00}.
\end{itemize}

We also estimate the rms deviations from the Hubble
  flow $\sigma_H$ for galaxies at different distances from the Local
  Group and find that in most of our model LV-candidates the rms
  peculiar velocities are consistent with observational values:
  $\sigma_H=50$\,km/s for distances less than 3\,Mpc and
  $\sigma_H=80$\,km/s for distances less than 8\,Mpc. At the distances
  4 (8) Mpc the observed overdensities of galaxies are 3.5-5.5
  (1.3-1.6) -- significantly larger than typically assumed.

\section*{Acknowledgments}
We thank I.D.\,Karachentsev for providing us an updated list of his
Catalog of Neighboring galaxies and D.I. Makarov for useful
discussions. We thank G. Yepes (UAM, Madrid), S. Gottl\"ober (AIP,
Potsdam), and Y. Hoffman (JHU, Jerusalem) for providing results of
their simulations and for  discussions.  A.\,Klypin acknowledges support of NSF grants to
NMSU.  Computer simulations used in this research were conducted on
the Columbia supercomputer at the NASA Advanced Supercomputing
Division and at the Leibniz-Rechenzentrum (LRZ), Munchen,
Germany. This work has been supported by the ASTROSIM network of the
European Science Foundation (ESF) (short visit grant 2089 of
A.~V. Tikhonov). A.~V. Tikhonov thanks the German Academic Exchange
Service for supporting his stay at the Astrophysical Institute Potsdam
in Autumn 2007 and Astronomy Department of NMSU for hosting in January
2008.

\label{lastpage}

\end{document}